\documentclass[useAMS,usenatbib]{mn2e}
\usepackage{graphics}
\usepackage{graphicx}
\usepackage{verbatim}

\title{High angular resolution imaging with stellar intensity interferometry using air Cherenkov telescope arrays}
\author[Paul D. Nu\~nez, Richard Holmes, David Kieda, Stephan LeBohec]{
Paul D. Nu\~nez$^{1}$\thanks{E-mail: pnunez@physics.utah.edu}, Richard Holmes$^{2}$, David Kieda$^{3}$, Stephan LeBohec$^{4}$\\
$^{1,3,4}$University of Utah, Dept. of Physics \& Astronomy, 115 South 1400 East, Salt Lake City, UT 84112-0830, USA\\
$^{2}$Nutronics Inc., 3357 Chasen Drive, Cameron Park, CA 95682, USA\\
}

\begin{document}
\maketitle

\begin{abstract}
  Optical stellar intensity interferometry with air Cherenkov telescope arrays, composed of nearly 100 telescopes, will
  provide means to measure fundamental stellar parameters and also open the possibility of model-independent
  imaging. In addition to sensitivity issues, a main limitation of image recovery in intensity 
  interferometry is the loss of phase of the complex degree of coherence during the measurement process. Nevertheless, 
  several model-independent phase reconstruction techniques have been developed. Here we 
  implement a Cauchy-Riemann based algorithm to recover images from simulated data.
  For bright stars ($m_v\sim 6$) and exposure times of a few hours, we find that scale features such as diameters, oblateness
  and overall shapes are reconstructed with uncertainties of a few percent. More complex images are also
  well reconstructed with high degrees of correlation with the pristine image. Results are further improved by using a forward algorithm.
\end{abstract}


\section{Introduction}
Even though Stellar Intensity Interferometry (SII) was abandoned in the 1970's,
there has been a recent interest in reviving this technique, mainly due to the
unprecedented $(u,v)$ plane coverage that future imaging air Cherenkov telescope (IACT) arrays will
provide \citep{cta, agis}. The possibility of probing stars at the 
sub-milliarcsecond scale and visible wavelengths has motivated new developments in instrumentation and
simulations, the latter being the focus of this paper.\\ 

Recent results obtained with amplitude (Michelson) interferometry have started to
reveal stars as extended objects (e.g. \citealt{coast, chara_review}), and with non-uniform light 
intensity distributions in the milliarcsecond scale. 
Such interesting results can be further investigated with SII taking advantage of the longer (km)
baselines and relative ease of observing at shorter (blue) wavelengths.
For example, measuring stellar diameters at different wavelengths, will make it possible to further investigate the
wavelength dependence of limb darkening,
\citep{Mozurkewich} and thus constrain stellar atmosphere models. Radii measurements with uncertainties of a few percent, along 
with spectroscopic measurements are necessary  
to constrain the position of stars in the HR diagram  (e.g. \citealt{Kervella}). With the
methods described in this paper, we show that diameters can in principle be measured
with accuracies better than $1\%$ when using realistic  array
configurations for future experiments such as CTA (Cherenkov Telescope
Array). As another example we can consider 
fast rotating B stars, which are ideal candidates for imaging oblateness,
pole brightening \citep{chara_review, von_zeipel}, radiatively driven mass loss \citep{Friend}, and perhaps even pulsation modes
\citep{pulsation}. The impact of rotation on stellar evolution is non-trivial, and 
several studies have been made in the subject (e.g. \citealt{martin, maeder}). Images of rotating stars have become available 
in the past few years (e.g. \citealt{altair, vega}),  and 
measurements of oblateness with accuracies of a few percent have been made. 
We will show that this is comparable to what 
can be achieved with SII using large arrays of Cherenkov telescopes. 
There is also the case of interacting binaries, for which we can measure angular separation,
diameters, and relative brightness. It may even be possible to measure mass transfer \citep{verhoelst}. Measurements of the angular separation
in binaries is crucial for determining the masses of stars. These masses must be found to within $\sim 2\%$ \citep{Andersen}
in order to test main sequence models. With the methods described in this paper, we show that angular 
separations can be found to within a few percent from reconstructed images. A more convenient and
accurate analysis (beyond the scope of this paper) does not require imaging, and should yield sub-percent uncertainties, 
so that testing main sequence models will become possible with SII.\\

In preparation for a large-scale SII observatory deployment, several 
laboratory experiments are in progress \citep{stephan.spie}. Their main goal is to measure light intensity correlation between two receivers. 
It is also worth mentioning the \emph{Star Base} observatory
(located in Grantsville, Utah) which consists of two $3\mathrm{m}$ light receivers separated by $24\,\mathrm{m}$ and which will be used
 to test high time resolution digital correlators,
and to measure the second order degree of coherence for a few stars. Various analog and digital correlator
technologies \citep{Dravins.timescale} are being implemented, and cross correlation of streams of photons with
nanosecond-scale resolution has already been achieved.\\

Intensity interferometry, unlike amplitude interferometry, relies on the
correlation between intensity fluctuations 
averaged over the spectral band at electronic (nanosecond) time resolution.
These averaged fluctuations are much slower than the (femtosecond) 
light wave period. This correlation is directly related to the 
complex degree of coherence $\gamma_{ij}$ as \citep{lipson}

\begin{equation}
|\gamma_{ij}|^2=\frac{<\Delta I_i \,\Delta I_j>}{<I_i> <I_j>}.
\end{equation}

Here, $<I_i>$ is the time average of the intensity received at a particular
telescope $i$, and $\Delta I_i$ is the intensity fluctuation. Measuring 
a second-order effect results in lower signal-to-noise ratio when compared to amplitude interferometry \citep{holder2}. This
sensitivity issue can be dealt with by using large light collection areas 
(such as those available with air Cherenkov telescopes), longer exposure times
and baseline redundancy. 
The low frequency fluctuation can be interpreted classically as the beat formed by
neighboring Fourier components. 
Since SII relies on low frequency
fluctuations, which are typically several orders of magnitude smaller than the
frequency of optical light,  it does not rely on the phase difference between light waves, 
but rather in the difference between the relative phases of the two components 
at the detectors \citep{Brown3}. The main advantage is the relative insensitivity to
atmospheric turbulence and the absence of a requirement for sub-wavelength precision
in the optics and delay lines \citep{Brown3}.
The complex mutual degree of coherence $\gamma$ is proportional to the Fourier
transform of the radiance distribution of the object in the sky (Van Cittert-Zernike theorem).
However, since with SII, the
squared-modulus of $\gamma$ is the measurable quantity, the main
disadvantage is that the phase of the Fourier transform is lost in the
measurement process. The loss of phase information poses a severe difficulty,
and images have in the past been reconstructed from the bispectrum technique, 
using monolithic apertures  (e.g. \citealt{lawrence}).
The imaging limitations
can be overcome using a model-independent phase recovery technique. Even though
several phase reconstruction techniques exist \citep{fienup}, we concentrate on a two
dimensional version of the one dimensional analysis introduced by \citet{Holmes}, which is based on the theory of analytic functions,
and in particular the Cauchy-Riemann equations.\\ 

Following recent successes in Gamma ray astronomy, a next generation Cherenkov telescope array is in a preparatory stage.
This project is currently known as CTA (Cherenkov telescope array) \citep{cta}, and will contain between 50 and 100 telescopes with 
apertures ranging between $5\,\mathrm{m}$ and $25\,\mathrm{m}$. In this paper we investigate the sensitivity 
and imaging capabilities of SII implemented on such an atmospheric Cherenkov 
telescope array. We start with a discussion of sensitivity (section \ref{sensitivity}), followed by a discussion of simulating 
noisy data as would be realistically obtained with such an array (section \ref{simulation}). Since data have
a finite sampling in the $(u,v)$ plane, we discuss our method of fitting an
analytic function to the data in order to estimate derivatives which are
needed for phase reconstruction (section \ref{fit}). We then briefly discuss
phase recovery (section \ref{phase_rec_sec})
and proceed to quantify the reconstruction quality using several criteria. We
start with the simple case of uniform disks (section \ref{disks}) and progressively increase the degree of
image complexity by including oblateness (section \ref{oblate_sec}), binary
stars (section \ref{binary_sec}), and  obscuring
disks and spots (section \ref{complex}).

\section{Sensitivity} \label{sensitivity}

The signal to noise ratio (SNR) for an intensity correlation measurement depends on the
degree of correlation $\gamma$, the area $A$ 
of each of the light receivers, the spectral density $n$ (number of photons per unit area per 
unit time, per frequency), the quantum efficiency $\alpha$, the electronic bandwidth $\Delta f$, 
and the observation time $t$. The SNR can be expressed as \citep{Brown3}

\begin{equation}
SNR=n(\lambda, T, m_v)\,A\;\alpha\;\gamma^2\sqrt{\Delta f t/2}. \label{snr}
\end{equation}

\begin{figure}
  \begin{center}
    \rotatebox{-90}{\includegraphics[scale=0.35]{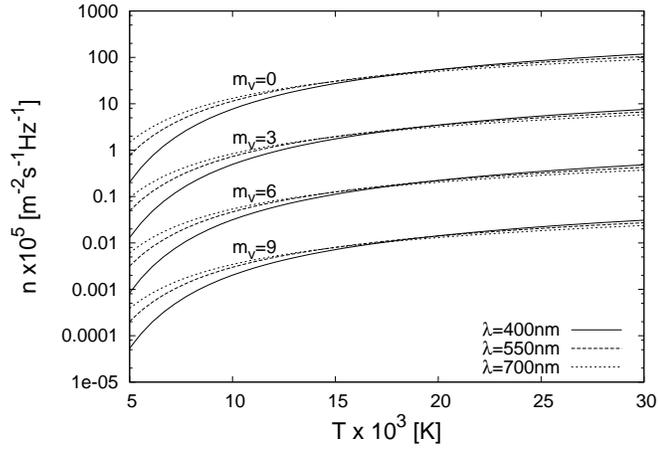}}
  \end{center}
  \vspace{0.5cm}
  \caption{\label{n_vs_T}Spectral density as a function of temperature for several different visual magnitudes and observed wavelengths. Atmospheric absorption as a function of the wavelength is not taken into account.}
\end{figure}

The spectral density $n$ is related to the visual magnitude $m_v$ of the star as well as its temperature $T$ 
and observing wavelength $\lambda$. The spectral density $n(\lambda, T, m_v)$ is the number of black body photons 
per unit area, per unit frequency and per unit time. The dependence of the visual magnitude $m_v$ is found by 
recalling that the flux for a $0^{th}$  magnitude star with a temperature of $9550^{\circ}\mathrm{K}$ observed at $550\,\mathrm{nm}$  
is $3.64\times 10^{-23}\mathrm{W m^{-2}Hz^{-1}}$ \citep{flux}. This in turn corresponds to a spectral density of
$10^{-4}\,\mathrm{m^{-2} s^{-1} Hz^{-1}}$. The 
spectral density as a function of temperature (for different visual
 magnitudes and observing wavelengths) is shown in Fig. \ref{n_vs_T}, and we see that at constant visual magnitude 
and observing wavelength, higher temperatures correspond to higher spectral densities. We find that the increase 
in temperature $\Delta T(\lambda, T, \Delta m_v)$ is approximately $\frac{\lambda k T^2}{h c}\Delta m_v$ for the 
range of temperatures and wavelengths considered in Fig. \ref{n_vs_T}. For example, at $400\,\mathrm{nm}$, a decrease of 1 visual 
magnitude is equivalent to increasing the temperature of the star from $5000\,\mathrm{K}$ to $5700\,\mathrm{K}$. For 
this reason high temperature objects are easier targets for SII.\\

\begin{figure}
  \begin{center}
    \rotatebox{-90}{\includegraphics[scale=0.45]{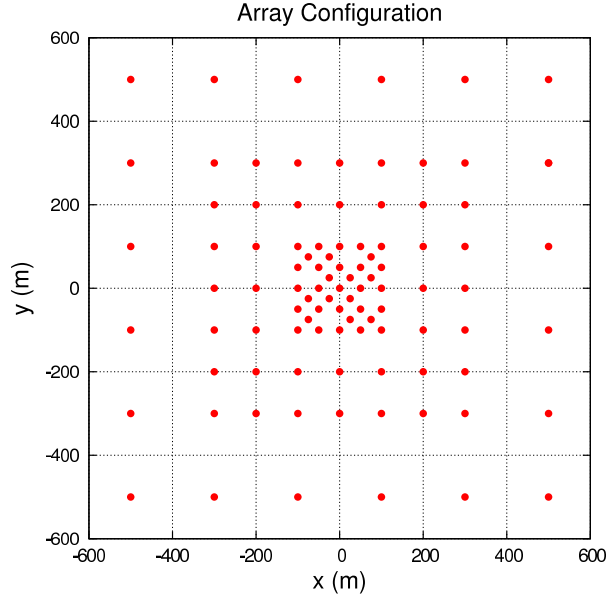}}
    \vspace{0.5cm}
  \end{center}
  \caption{\label{array} Array configuration used for our analysis.}
\end{figure}

\begin{figure}
\begin{center}
  \rotatebox{-90}{\includegraphics[scale=0.5]{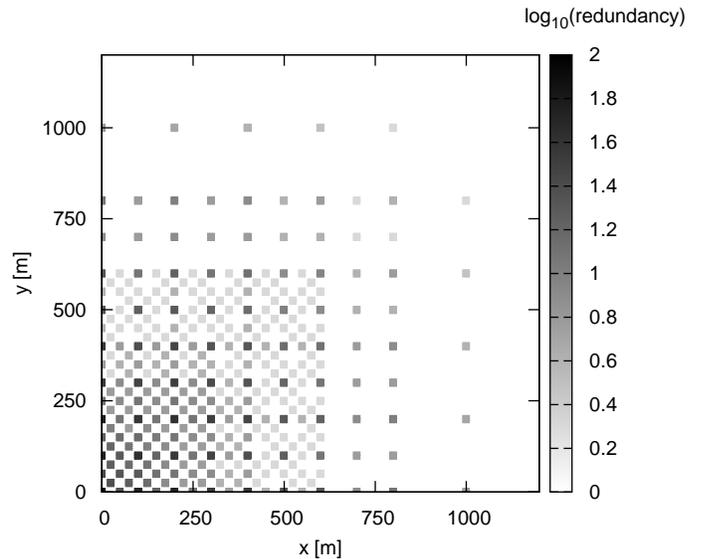}}
\end{center}
\caption{\label{baselines} A total of 4656 non-zero baselines are available in the array design used in 
this study. Gray scale measures the $log$ of the baseline redundancy. Since the array shown in 
Fig. \ref{array} is almost symmetric with respect to $x$ and $y$, only a quadrant of the $(u,v)$ plane is displayed.}
\end{figure}

We use a preliminary design of the CTA project as an array configuration \citep{cta_simulation}, which is shown in
Fig. \ref{array}. This array contains $N=97$
telescopes and $N(N-1)/2=4646$ baselines (many of which are redundant) which are shown in Fig. \ref{baselines}. Each detector is
assumed to have a light collecting area of $100\,\mathrm{m}^2$  and a
light detection quantum efficiency of $\alpha=0.3$. Using a $\lambda/D$ criterion, we find that 
the largest baselines of $1.5\,\mathrm{km}$ resolve angular
scales of $\sim 0.05\,\mathrm{mas}$ at $400\,\mathrm{nm}$.
The smallest $48$  baselines of $35\,\mathrm{m}$ resolve angular scales 
of $\sim 2\,\mathrm{mas}$ . However, we show in section \ref{disks}, 
that the largest angular scales that can be realistically \emph{imaged} with our analysis, 
in a model independent way, are more determined by 
 baselines of $\sim 70\,\mathrm{m}$. This is because the estimation
of derivatives of the phase (needed for phase recovery) degrades as the number of 
baselines is reduced. Baselines of $70\,\mathrm{m}$ resolve 
angular scales of $\sim 1.2\,\mathrm{mas}$ at $400\,\mathrm{nm}$.\\

\begin{figure}
  \begin{center}
    \rotatebox{-90}{\includegraphics[scale=0.35]{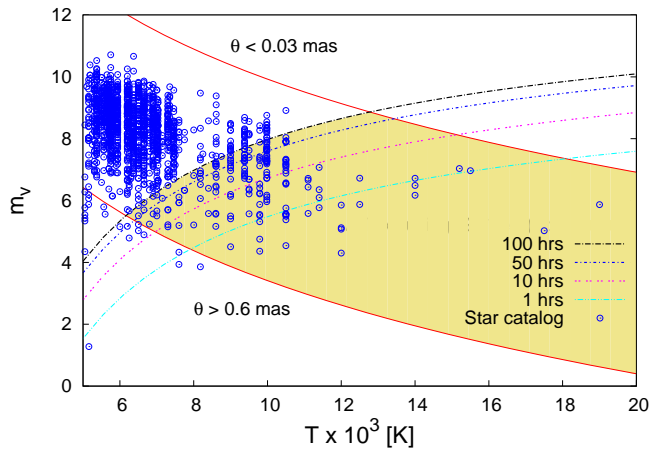}}
  \end{center}
  \vspace{0.5cm}
  \caption{\label{MvsT} The four parallel curves indicate the maximum detectable visual magnitude 
as a function of the temperature for several exposure times. Each of these four curves corresponds to
5 standard deviation measurements and $|\gamma|=0.5$. Also shown is the (shaded) region corresponding
to angular radii between $0.03\,\mathrm{mas}$ and $0.6\,\mathrm{mas}$, and observable within 
$100\,\mathrm{hrs}$ of observation time. The positions in $(T,m_v)$ space of 2000 stars from the 
JMMC stellar diameters catalog \citep{catalog} are included. 
 }
\end{figure}

These order-of-magnitude considerations are taken into account when performing
simulations and image reconstructions, i.e. the minimum and maximum size of
pristine images that can be reconstructed by data analysis, do not go far beyond these limits. More precise array resolution 
limits are presented
in section \ref{disks} (diameters ranging between $0.06\,\mathrm{mas}\,-\,1.2\,\mathrm{mas}$) . By combining 
these angular scales with the SNR (eq. \ref{snr}),  
we obtain Fig. \ref{MvsT}. This figure shows the highest visual magnitude, for which photon correlations (with $|\gamma|=0.5$)
can be detected (5 standard deviations), as a function of the temperature, and for several different exposure times. Also 
shown in Fig. \ref{MvsT}, is the shaded region 
corresponding to angular diameters between $0.03\,\mathrm{mas}$ and $0.6\,\mathrm{mas}$ \footnote{These curves of constant angular size can be found approximately by recalling that the visual magnitude $m_v$ is related to a calibrator star of visual magnitude $m_0$ by: $(m_v-m_0)=-2.5\log{F/F_0}$. Here, $F$ and $F_0$ refer to the flux in the visual band . To express $m_v-m_0$ as a function of the angular
size, note that flux is proportional to $\theta^2T^4$, where $\theta$ is the angular size and $T$ is the temperature of the star.}, and observable 
within $100\,\mathrm{hrs}$. From the Figure we 
can see how correlations of photons from faint stars can be more easily detected if they are hot. To quantify the number of stars
for which photon correlations can be detected with the IACT array, we perform use the JMMC stellar diameters catalog \citep{catalog}. We 
find that $\sim 1000$ (out of $\sim33000$) stars from the JMMC catalog can be detected 
within $1\,\mathrm{hr}$, correlations from $\sim 2500$ stars can be detected within 
$10\,\mathrm{hrs}$, and $\sim 4300$ can be detected within $50\,\mathrm{hrs}$. In Fig. \ref{array},
we show a random sample of 2000 stars (out of $\sim33000$) 
from the JMMC catalog. Interstellar reddening may play a role in reducing the number of measurable targets

\section{Simulation of realistic data} \label{simulation}

The original ``pristine'' image consists of $2048\times2048$
pixels corresponding to $\sim10\,\mathrm{mas}\times10\,\mathrm{mas}$ of angular extension and a
wavelength of $\lambda= 400\,\mathrm{nm}$. The
Fourier transform of the image is then performed via an FFT algorithm and
normalized so that its value is one at zero baseline. This results in a
Fourier transform sampled every $\mathrm{\sim(8\,\mathrm{m})/\lambda}$, 
i.e. $2\times 10^7$ cycles per radian field-of-view at a wavelength $\lambda$ of $400\,\mathrm{nm}$. We then find 
the squared-modulus of the degree of coherence between the members of each pair of telescopes.
This is obtained from a linear interpolation of the squared Fourier magnitude in the finely
sampled FFT. Diurnal motion is not taken into account in the simulations. Diurnal motion plays 
a significant role in increasing the $(u,v)$ coverage when exposure times are long. 
As a consequence there is less $(u,v)$ coverage in the simulations since 
projected baselines do not drift with time. The smaller
$(u,v)$ coverage is however compensated by smaller statistical error in the correlation measurements.\\

\begin{figure}
  \begin{center}
    \includegraphics[scale=0.35, angle=-90]{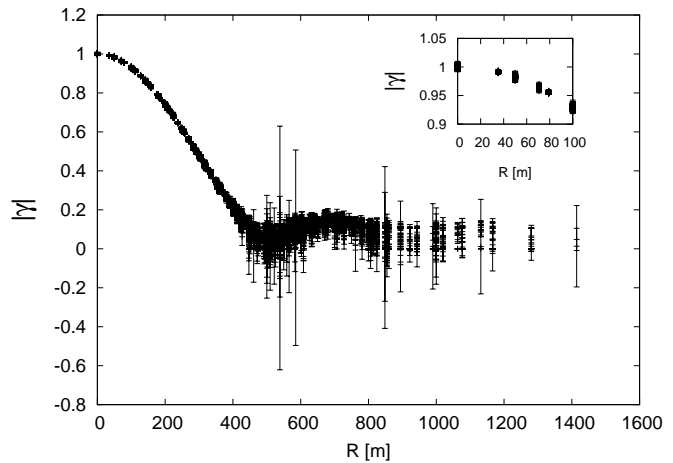}
  \end{center}
  \vspace{0.5cm}
  \caption{\label{example_simulation} Example of simulated data for a $3^{rd}$ magnitude 
uniform disk star ($T=6000^{\circ}K$) of radius $0.1 \,\mathrm{mas}$ and $10\,\mathrm{hrs}$ of observation time. Here we show
$|\gamma|$ (instead of the directly measured $|\gamma|^2$) as a function of telescope separation.}
\end{figure}

The final step in the simulation phase is the addition of noise to the
correlation at each baseline. This noise was found to be
Gaussian by performing the time integrated product of two random streams of
simulated photons as detected by a pair of photo-multiplier tubes. The standard
deviation of the noise added to each pair of telescopes is calculated
from eq. \ref{snr}. In this paper we take the signal bandwidth to be $\Delta
f=200\,\mathrm{MHz}$. An example of simulated data as a function of telescope separation 
is shown in Fig. \ref{example_simulation}. This corresponds to a $3^{rd}$ magnitude 
uniform disk star ($T=6000^{\circ}\mathrm{K}$) of radius $0.1 \,\mathrm{mas}$ 
and $10\,\mathrm{hrs}$ of observation time. The software used for the simulations, as well as the 
analysis\footnote{See sections \ref{fit} and \ref{phase_rec_sec} for details on the analysis. Some 
variants of the analysis software were developed in \textbf{MATLAB}. All software is available upon request.}, was developed in \textbf{C}.\\ 

\section{Fitting an Analytic function to the data} \label{fit}

The estimation of derivatives of the Fourier log-magnitude is at the heart of the Cauchy-Riemann phase
recovery algorithm (section \ref{phase_rec_sec}), and is thus greatly simplified when data is known on
a square grid rather than in a `randomly' sampled way as is directly available
from observations. Once simulated data are available (or observations in the future), an analytic function is fitted to
the data.\\

An analytic function can be expressed as a linear combination of basis functions. When data
$f(x_i)\equiv|\gamma(x_i)|^2$ are available at baselines $x_i$, with uncertainty
$\delta f(x_i)$, the coefficients of the basis functions can be found by minimizing the
following $\chi^2$:

\begin{equation}
\chi^2=\sum_i\left[\frac{\left(f(x_i)-\sum_ka_kg_k(\alpha R(x_i))\right)}{\delta f(x_i)}\right]^2. \label{chi2}
\end{equation}

Each $a_k$ is the coefficient of a basis function $g_k$.  The constant $\alpha$ is a scaling factor,
 and $R$ is a rotation operator. The scaling factor and rotation angle are found by first performing a two-dimensional 
Gaussian fit. Finding the appropriate scale and rotation angle has the advantage of reducing the number of
basis elements needed to fit the data.\\

Basis functions which tend to zero
at very large baselines, where data are scarce (see Fig. \ref{baselines}), are ideal. For this reason, we use 
Hermite functions. There are situations where data are more easily fit with a different set of basis functions, e.g. 
a binary with unresolved members. In such a situation, data do not rapidly tend to zero at large baselines, so 
the Hermite function fit may contain a large number of elements and result
in high frequency noise where data are scarce\footnote{A basis set consisting of products of sines and cosines is more
suitable in this situation}. The best choice of basis functions may therefore depend on the structure of the object. However, 
for consistency, we use the Hermite fit for all the objects that we analyze,
and find that it gives reasonably good results.\\

The $\chi^2$ minimization problem can be turned into a linear system by setting the set of partial
derivatives $\frac{\partial \chi^2}{\partial a_k}$ to zero. We typically start
with a small number of basis elements, say eight, and only increase the number
of basis elements if the optimized reduced $\chi^2$ is greater than some predefined value. 

\section{Cauchy-Riemann phase reconstruction} \label{phase_rec_sec}

In order to perform a model-independent image reconstruction, 
the phase of the Fourier transform needs to be recovered from
magnitude information only \citep{lipson}. Since the image is real, the Fourier magnitude 
is symmetric with respect to the origin in the $(u,v)$ plane. Lack of phase information results 
in the reconstructed image being arbitrary up to a translation and reflection. We discuss 
 the one-dimensional phase reconstruction prior to developing 
the two-dimensional analysis. We can start by first approximating the 
continuous Fourier transform $I(x)$ by a discrete one, i.e. $I(m\Delta x)=\sum_j
\mathcal{O}(j \Delta \theta) e^{ijmk_0 \Delta x \Delta \theta}$, where
$\mathcal{O}(\theta)$ is the image in the sky and $k_0$ is the usual wave
vector.  Then it becomes convenient to express the discrete Fourier transform in phasor
representation , i.e. $I(z)=R(z) e^{i\Phi(z)}$ where $z\equiv e^{imk_0 \Delta x
  \Delta \theta}$ is complex. Since the discrete Fourier transform is a
polynomial in $z$, the theory of analytic functions can be applied. As a consequence, we obtain
the Cauchy-Riemann equations in polar coordinates, which relate the
log-magnitude and the phase along the purely real and imaginary axis. If data
were available along the purely real or purely imaginary axis, we could directly solve for the
phase by integrating along these directions. However, since $z$, the
independent variable of the Fourier transform ($z\equiv e^{imk_0 \Delta x
  \Delta \theta}$), has modulus equal to 1, the phase differences that we seek
lie along the unit circle in the complex plane. On the other hand, one can
show that by using the Cauchy-Riemann equations, the phase differences
along the radial direction in the complex plane\footnote{If $\xi$ 
is the real axis and $\psi$ is the imaginary axis, then a difference 
along the radial direction is $\Delta\xi+i\Delta\psi$ (assuming $\Delta\xi=a\xi$ and $\Delta\psi=a\psi$, 
where $a$ is a proportionality constant).} are directly related to the differences in the logarithm of 
the magnitude which is available from the data (see appendix A for more details).\\

The procedure to find the phase consists of first assuming a general solution form, then taking 
differences in the radial direction of the complex z-plane, and finally fitting the data to the radial differences of
the assumed solution. A general form of the phase  
can be postulated by noting that the phase is a solution of the Laplace equation in the 
complex plane (applying the Laplacian operator on the phase and using the Cauchy-Riemann equations 
yields zero). Since the phase differences
are known along the radial direction in the complex plane we can take radial differences of the 
general solution and then fit the log-magnitude differences (available from the data) to
the radial differences of the general solution.\\

\begin{figure}
  \begin{center}
    \includegraphics[scale=0.4]{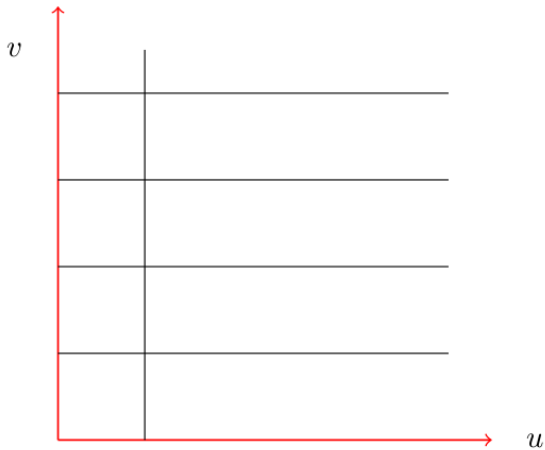}
    \caption{\label{slices} Schematic representation of two-dimensional phase reconstruction approach. Several parallel slices are related to
      a single orthogonal slice.}
  \end{center}
\end{figure}

We can think of this one-dimensional reconstruction as a phase estimation along a single slice in the Fourier plane. A 
generalization to two dimensions can be made by doing the same procedure for several slices as 
described in Fig. \ref{slices}. In fact, the requirement that a two-dimensional complex function $(z_x, z_y)$ be analytic,
is equivalent to satisfying the Cauchy-Riemann equations in both $z_x$ and $z_y$ \citep{2-d_complex}. The direction of the slices 
is arbitrary, however for simplicity we reconstruct the phase along an
arbitrary set of perpendicular directions in the Fourier plane, 
and noting that one can relate all slices through a single orthogonal slice, i.e. once the phase at the origin is set to zero, 
the single orthogonal slice sets the initial values for the rest of the slices.\\

 One can also require that the phase at a particular point 
in the complex plane be exactly equal when reconstructed along $z_x$ or $z_y$ since each reconstruction is arbitrary up to a constant (piston) and a
linear term (tip/tilt). However, imposing this requirement results in a severely over-determined linear system. More precisely, by imposing equality 
in $n^2$ points in the complex plane, and having $2n$ slices (each with an unknown constant and linear term), results in a linear system of $n^2$ equations
and $4n$ unknowns. Alternative methods of requiring slice consistency are a possible way of improving phase reconstruction, but are beyond 
the scope of this paper.\\

 The Cauchy-Riemman approach, with horizontal or vertical slices, and a single orthogonal slice, 
gives reasonably good results,  however, it is not the only
possible approach.  We have also investigated Gerchberg-Saxton phase
retrieval, Generalized Expectation Maximization, and other variants of the
Cauchy-Riemann approach. It is premature to conclude which of these approaches 
is best at this time, given the limited imagery and SNR levels that have been explored. However, the 
Cauchy-Riemman approach has shown to give better results in a number of cases \citep{Holmes.spie}.

\section{Imaging capabilities}

We investigated the imaging capabilities for simple objects\footnote{For preliminary study see \citet{nunez.spie}.}, namely uniform
disk-like stars, oblate rotating stars, binaries, and more complex
images. For most of the objects that we consider, image reconstruction is 
not necessary, i.e. from the Fourier magnitude alone, one can extract radii, 
oblateness, relative brightness in binaries, etc. Estimation of these 
parameters is probably more accurate when extracted directly from Fourier
 magnitude data only, especially if some a-priori knowledge of the original 
image is available. However, measuring simple parameters from reconstructed 
images is the first step in quantifying reconstruction capabilities with IACT 
arrays. We assume no a-priori knowledge of the images that are being 
reconstructed, and then do a statistical study of the uncertainties of the reconstructed parameters.

\subsection{Uniform disks \label{disks}}

\begin{figure}
  \begin{center}
    \rotatebox{-90}{\scalebox{0.4}{\includegraphics{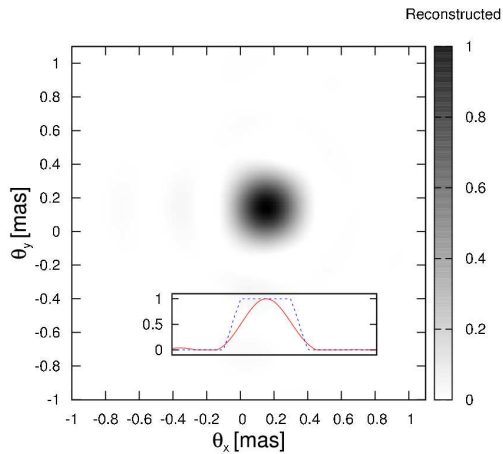}}}
  \end{center}
  \caption{\label{radius_a} Example of a reconstructed uniform disk of radius $0.1\,\mathrm{mas}$. 
    Also shown is a slice of the reconstructed  image (solid line) compared to 
    a slice of the pristine image convolved with the PSF of the array (dashed line).}
\end{figure}

\begin{figure}
  \begin{center}
    \rotatebox{-90}{\scalebox{0.35}{\includegraphics{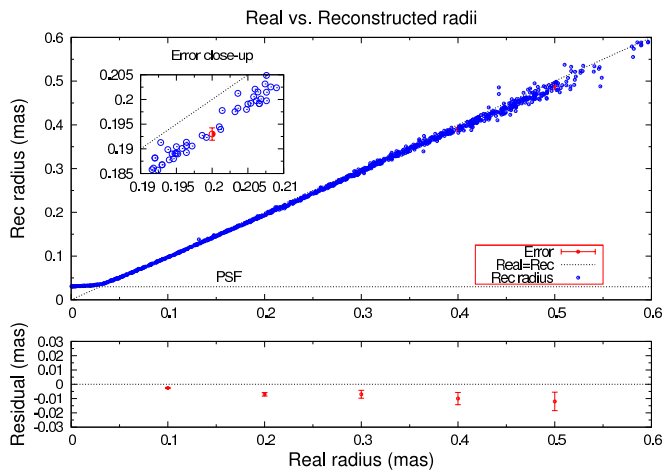}}}
  \end{center}
  \vspace{0.5cm}
  \caption{\label{radius_b}  Simulated vs. Reconstructed radii for magnitude 6 stars 
with 50 hours of observation time (see footnote \ref{long_exposures}). The 
top sub-figure shows the uncertainty for a $0.2\,\mathrm{mas}$ measurement. The 
bottom sub-figure shows the residual (Reconstructed-Real) along with the uncertainty in the radius.}
\end{figure}

\begin{figure}
  \begin{center}
    \rotatebox{-90}{\scalebox{0.32}{\includegraphics{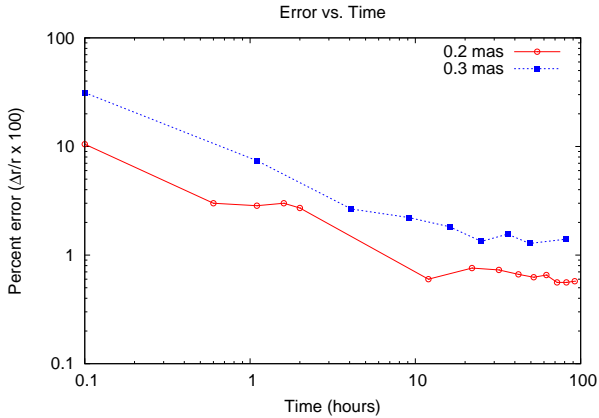}}} 
  \end{center}
  \vspace{0.5cm}
  \caption{\label{uncertainty_b} Percent error as a function of time for two reconstructed radii. This 
    error was found by performing several reconstructions for each  radius and exposure time, 
    and then taking the standard deviation of the reconstructed radius.}

\end{figure}

In order to quantify the uncertainty in the reconstructed radius, we 
simulate data corresponding to $6^{th}$ magnitude stars ($T=6000^{\circ}\,\mathrm{K}$) with disk radii up to
$1\,\mathrm{mas}$ for 50 hours of exposure time\footnote{\label{long_exposures} This brightness and exposure time 
correspond to uncertainties in the 
simulated data of a few percent. Such long exposure times can be reduced to 
a few hours as is shown in Fig. \ref{uncertainty_b}}. An example of such a reconstruction is shown in Fig.
\ref{radius_a}, where the radiance is shown in arbitrary units between 0 and
1. For a uniform disk, the reconstructed phase is null in the first lobe, and we find that the rms deviations from the true
phase are approximately $0.19\,\mathrm{rad}$ in the null zone.  A first look at the reconstruction in Fig. \ref{radius_a} reveals
that the edge of the reconstructed disk is not sharp, so a threshold in the
radiance was applied for measuring the radius. The radius is measured
by counting pixels above a threshold and noting that the area of the disk is
proportional to the number of pixels passing the threshold. After experimenting
with different radii, we chose the threshold for measuring the radius to be
0.2. We can now compare the simulated and
reconstructed radii as is shown in Fig. \ref{radius_b}, where each point in the
Fig. corresponds to an individual simulation (including noise) and
reconstruction. Further optimization in the threshold for measuring the radius 
should yield a slope even closer to unity in Fig. \ref{radius_b}.\\

\begin{figure}
  \begin{center}
    \rotatebox{-90}{\scalebox{0.32}{\includegraphics{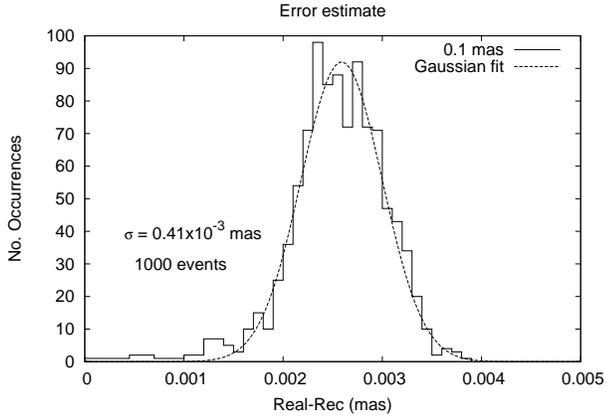}}}
  \end{center}
  \vspace{0.5cm}
  \caption{\label{uncertainty_a} Histogram of real radius minus reconstructed
  radius for 50 hours of exposure time on a $6^{th}$ magnitude star ($T=6000\,\mathrm{K}$) of $0.1\,\mathrm{mas}$ radius. }
\end{figure}

Fig. \ref{radius_b} clearly shows that stellar radii ranging from $0.03\,\mathrm{mas}$
to $0.6\,\mathrm{mas}$ can be measured with uncertainties ranging between sub-percent and 
a few percent (Figures \ref{uncertainty_a} and \ref{uncertainty_b}). It can 
be seen from Fig. \ref{radius_b}, that the
uncertainty increases linearly as a function of the pristine (simulated) radius. This
is due to a decrease in the number of baselines that measure a high degree of correlation
 when the angular diameter increases. As the pristine radius $\theta$
decreases, the distance to the first zero in the correlation increases as
$\sim \theta^{-1}$, so the number of telescopes contained within the airy
disk increases as $\sim \theta^{-2}$. Consequently, decreasing the pristine radius is
equivalent to increasing the number of independent measurements by a factor of
$\sim \theta^{-2}$. Since the uncertainty decreases as the square root of the number of independent
measurements, the error decreases linearly with the radius. For radii above
$0.6\,\mathrm{mas}$, there are simply not enough baselines to constrain the Fourier
plane information for image reconstruction. For radii greater than $0.6\,\mathrm{mas}$,
the distance to the first zero in the degree of correlation is of the order of $100\,\mathrm{m}$, 
but only baselines at $35\,\mathrm{m}$ and $50\,\mathrm{m}$ are 
capable of measuring the Fourier magnitude with more than 3 standard deviations (see eq. \ref{snr}). In 
Fig. \ref{uncertainty_b} we show the relative percent error (RMS of a statistic) 
as a function of time for two different radii, where it can be seen that a
 relative error of a few percent is achieved after only a few hours.

\subsection{Oblate stars\label{oblate_sec}}

\begin{figure}
  \begin{center}
    \rotatebox{-90}{\includegraphics[scale=0.39]{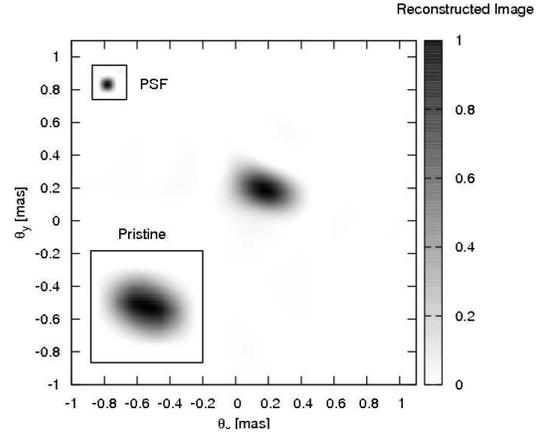}}
  \end{center}
  \caption{\label{oblate_example_a} Simulated and reconstructed oblate rotator of magnitude 6 and 50 hours
  of observation time.}
\end{figure}

For oblate stars we use the same magnitude and exposure
parameters that are used for disk-like stars. Uniform oblate stars can be
described by three parameters: the semi-major axis $a$, the
semi-minor axis $b$, and the inclination angle $\theta$. Judging from the
limitations obtained from reconstructing disks, we produce 
pristine images whose values for $a$ and $b$ are random numbers less than $1\,\mathrm{mas}$, and choose $1\leq a/b \leq2$ . The value
of the inclination angle $\theta$ also varies randomly between $0^\circ$ and
$90^\circ$. A typical image reconstruction can be seen in Fig.
\ref{oblate_example_a}.\\

\begin{figure}
  \begin{center}
    \rotatebox{-90}{\includegraphics[scale=0.32]{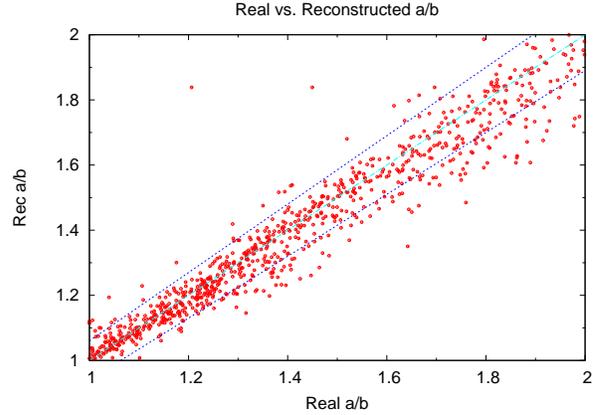}}
  \end{center}
  \vspace{0.5cm}
  \caption{\label{oblate_example_b} Real vs. reconstructed $a/b$ for oblate
  stars. The distance between the two linear envelopes is 2
  standard deviations.  All pristine images
  that have either $a>0.5\,\mathrm{mas}$ or $b>0.4\,\mathrm{mas}$ are not included}
\end{figure}

After applying a threshold on pixel values as was done for disk shaped stars,
the reconstructed parameters are found by calculating the inertia
tensor of the reconstructed image. The eigenvalues and
eigenvectors of the inertia tensor provide information for the reconstructed values of $a$, $b$ and
$\theta$. To do this, we make use of the relation between the matrix eigenvalue and the semi-major/minor axes
$I_{xx}=\frac{1}{4}a^2M$, where $M$ is the integrated brightness. A similar relation for $I_{yy}$ holds
for the semi-minor axis $b$. \\ 

The resulting reconstructed semi-major/minor axes as a function of their real values have a similar
 structure as the scatter plot for reconstructed radii shown in Fig.
\ref{radius_b}, and are well reconstructed up to $0.5\,\mathrm{mas}$ within a few percent. 
In Fig. \ref{oblate_example_b}, it can be seen how the uncertainty in the reconstructed oblateness $a/b$
increases with increasing oblateness. As with disk shaped stars (section \ref{disks}), 
the uncertainty in the reconstructed semi-major/minor axes decreases as the square root 
of the number of baselines measuring a high degree of correlation. Therefore, the 
uncertainty in the reconstructed semi-major/minor axes is proportional to $\sim \sqrt{ab}$, and the error in the reconstructed oblateness 
is proportional to $\sim \sqrt{a/b+a^3/b^3}$.\\

\begin{figure}
  \begin{center}
    \rotatebox{-90}{\includegraphics[scale=0.32]{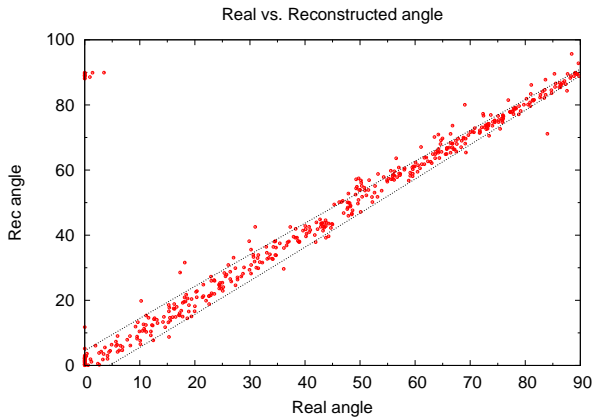}}
  \end{center}
  \vspace{0.5cm}
  \caption{\label{theta_oblate} Real vs. reconstructed inclination angle for oblate stars. All pristine images
    that have either $a>0.5\,\mathrm{mas}$, $b>0.4\,\mathrm{mas}$ or $a/b<1.1$ are not included.  
    We also note that reconstructed angles are always less than $45^\circ$
    due to the fact that the Fourier magnitude data is the same for pristine
    images flipped about the x, y or x and y axes. Therefore, for all 
    pristine angles larger than $45^\circ$, we replace the reconstructed angle by
    $\theta_{rec}'=90^\circ-\theta_{rec}$. }
\end{figure}

The reconstructed inclination angle as a function of the pristine angle is shown in Fig. \ref{theta_oblate}, and
several factors play a role in the uncertainty of the reconstructed value. For a fixed value of $a$ and $b$, 
the orientation of the telescope array with respect to the main lobe of the Fourier magnitude 
determines the number of baselines that measure a high degree of correlation. The number of 
baselines that measure a high degree of correlation is greater when the main lobe of the Fourier 
magnitude is aligned with the $x$ or $y$ direction of the array (see Fig. \ref{array}), and is smaller by a factor of $\sim \sqrt{2}$ (assuming a uniform grid of telescopes)
when its main axis is at $45^{\circ}$ with respect to the array. However, the uncertainty (proportional to spread of points) 
in Fig. \ref{theta_oblate} does not appear to be symmetric at $0^{\circ}$ and $90^{\circ}$, and is smaller at $90^{\circ}$. This due to the
way the phase is reconstructed, i.e. due to the slicing of the Fourier plane 
along the $u$ or $v$ directions (see section \ref{phase_rec_sec}). In the case of Fig. \ref{theta_oblate}, the $(u,v)$ plane is
sliced along the $u$ direction, with a single orthogonal reference slice along the $v$ direction. The 
main lobe of the Fourier magnitude  
of an oblate star has more slices passing through it when it is elongated along the $v$ 
direction (corresponding to an inclination angle of $90^{\circ}$ in image space), yielding a better reconstruction. This is in contrast to
the orthogonal case of $0^{\circ}$, where the main lobe of the Fourier magnitude has a smaller number of slices passing through it.

\subsection{Binary stars\label{binary_sec}}

\begin{figure}
  \begin{center}
    \rotatebox{-90}{\includegraphics[scale=0.4]{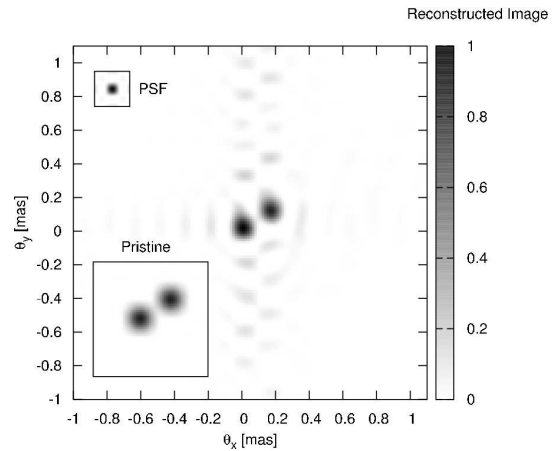}} 
  \end{center}
  \caption{\label{binary_example_a} Simulated and reconstructed binary of magnitude 6
    and 50 hours of observation time.} 
\end{figure}

Simulated data are generated for $5^{th}$ magnitude binary stars ($T=6000^{\circ}\,\mathrm{K}$), and an exposure of 
50 hours after noting that the uncertainty in the degree of correlation is of the order of a few percent (eq. \ref{snr}).
Binaries stars are parameterized by the radii $r_1$ and $r_2$ of each star, 
their separation $d$, position angle $\theta$, and relative brightness in arbitrary units between 0 and 1. 
We generate pristine images with random parameters within
the following ranges: radii are less than $0.3\,\mathrm{mas}$, angular separations are
less than $1.5\,\mathrm{mas}$, the relative brightness per unit area is less than or equal to 1, and
the orientation angle is less than $90^\circ$. A typical reconstruction can
be seen in Fig. \ref{binary_example_a}.\\

\begin{figure}
  \begin{center}
    \rotatebox{-90}{\includegraphics[scale=0.32]{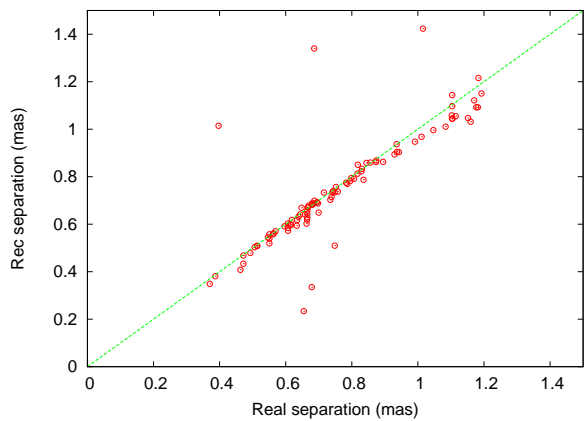}}
  \end{center}
  \vspace{0.5cm}
  \caption{\label{binary_example_b} Real vs. reconstructed angular separation in
    binary stars. Binary stars whose relative brightness is less than 0.3 are not included in this plot. }
\end{figure}

To measure the reconstructed parameters we identify the two brightest spots whose pixel values exceed 
a threshold of 0.2. We then find the radius for each bright spot and its centroid position.
 Identifying spots is a non trivial task
in noisy reconstructed images and our analysis sometimes fails to identify
the ``correct'' reconstructed spots. For example, a common issue is that
close reconstructed spots that are faintly connected by artifacts, are sometimes 
interpreted as a single spot. It should be again stressed that image reconstruction may not be
the best way to measure reconstructed parameters. For example, the data can just as well be
fit by the general form of the Fourier magnitude of a resolved binary system (containing only a few parameters).\\

In Fig. \ref{binary_example_b}, we show reconstructed angular separations
as a function of their real values. The reconstructed values of the angular separation are found
to within $\sim 5\%$ of their real values and  cannot be much 
larger than what is allowed by the smallest baselines. We find that stars separated 
by more than $d_{max}\approx 0.75\,\mathrm{mas}$ are not well reconstructed since the variations in the Fourier magnitude 
start to become comparable to the shortest baseline.\\

\begin{figure}
  \begin{center}
    \rotatebox{-90}{\includegraphics[scale=0.32]{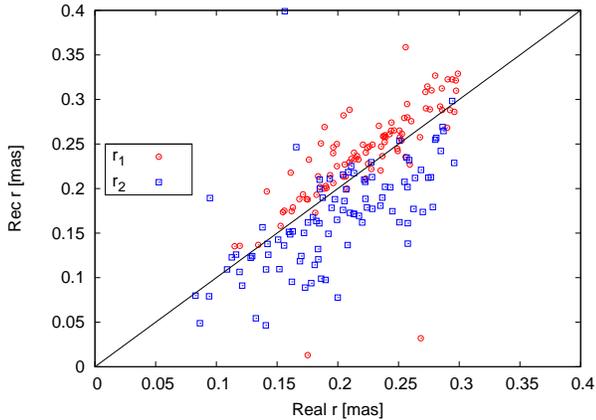}}
  \end{center}
  \vspace{0.5cm}
  \caption{\label{relative} Real versus reconstructed
    radii for binary stars. We only include cases where one of the members is less than 3 times brighter than the other.}
\end{figure}

In Fig. \ref{relative} we show the reconstructed values of the radii as a function of their pristine 
values. We find $\sim 10\%$ uncertainties in each of the reconstructed radii. 
Aside from the angular separation, a variable that plays a role 
in successfully reconstructing pristine radii is the 
ratio of absolute brightness\footnote{Product 
of relative brightness per unit area (in arbitrary linear units between 0 and 1) and relative area of both stars} of both binary members. When 
one of the two members is more than $\sim 3$ times brighter than the other, the fainter star
is found to be smaller than its pristine value, and sometimes not found at all when 
one of the members is more than $\sim 10$ times brighter than the other. This is in part because the sinusoidal variations
in the Fourier magnitude start to become comparable to the uncertainty. For example: a non-resolved binary star
with one component 20 times brighter that the other, has relative variations of $\sim 10\%$. With all the redundant baselines,
 a few percent uncertainty in the measured degree of correlation is sufficient to accurately measure these variations. 
However, when the binary components 
are resolved, the relative variations decrease with increasing baseline and baseline redundancy is not sufficient to 
reduce the uncertainty in the measurement of the Fourier magnitude. This signal to noise issue can of course be
improved by increasing exposure time.\\

There are also issues related to algorithm performance. One such problem has to do with the fit of the
data to an analytic function (see section \ref{fit}). When the scale of the fit (found by an initial Gauss fit)
is found to be too small, too many basis elements are
used to reconstruct the data, and high frequency artifacts appear in reconstructions. Small initial scales are
typically related to the binary separation as opposed to the size of individual components, and it is the latter which 
correctly sets the scale of the fit. Artifacts may be then mistaken for binary 
components, and incorrect reconstructed parameters may be found. Results improve significantly when either
the correct scale is set or (model-independent) image post-processing is performed (see section \ref{mira}).

\subsection{Featured images\label{complex}}

\begin{figure}
  \begin{center}
    \rotatebox{-90}{\includegraphics[scale=0.4]{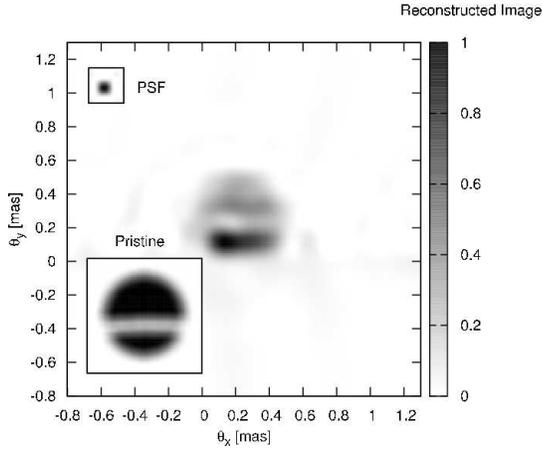}} 
  \end{center}
  \caption{\label{horizontal_disk} Star with obscuring disk (raw reconstruction). This corresponds to $4^{th}$ magnitude and 10 hrs of observation time. The correlation between the real and reconstructed image is $c=0.947\pm 0.001$. Note that an inverted gray scale is used.
}
\end{figure}

We now present two examples of more complex images, and show that the capabilities can, to a large extent, be understood from results of 
less complex images, such as uniform disks and binaries. In Fig. \ref{horizontal_disk} we show the reconstruction of a star with a dark 
band (obscuring disk), corresponding to a $4^{th}$ magnitude star and 10 hrs of observation time. The metric used to quantify the 
agreement with the pristine image (bottom left corner of Fig. \ref{horizontal_disk}) is a normalized correlation\footnote{For two images $A_{i,j}$ and $B_{i,j}$, the normalized correlation $C_{i,j}$ is $C_{i,j}=Max_{k,l}\left\{N^{-2}(\sigma_A\sigma_B)^{-1} \sum_{i,j}^{N, N}A_{i,j}B_{i+k, j+l}\right\}$, where $\sigma_A$ and $\sigma_B$ are the standard deviations of images $A$ and $B$. } 
whose absolute value ranges between 0 (no correlation) and 1 (perfect correlation/anti-correlation). To quantify the uncertainty in the correlation, we perform the noisy simulation and reconstruction several times, and find the standard deviation of the degree of correlation. In the case of Fig. \ref{horizontal_disk}, the correlation $c$ is $c=0.947\pm 0.001$. Note that the uncertainty in the correlation is apparently small, and the image reconstruction is not perfect, which implies that the reconstruction is not only affected by the SNR level, but also by the reconstruction algorithm performance.\\

 In order to determine the confidence with which we can detect the feature (dark region within the disk), this correlation is compared to the correlation of the reconstruction, and a featureless image, i.e. a uniform disk whose radius matches the radius of the pristine image. This comparison allows us to quantify the confidence with which we actually detect the feature (dark spot). Consequently, the correlation of the reconstructed image with a uniform disk  is $c=0.880\pm 0.001$, which is lower by 61 standard deviations.\\

\begin{figure}
  \begin{center}
    \rotatebox{-90}{\includegraphics[scale=0.4]{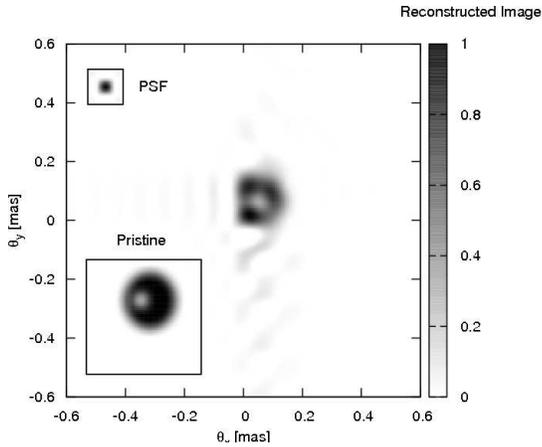}}
  \end{center}
  \caption{\label{dark_spot} Star with dark spot (raw reconstruction). This corresponds to 
    a $4^{th}$ magnitude star, 10 hrs of observation time and a degree of correlation of $0.940\pm0.001$.}  
\end{figure}

Another example of a complex image reconstruction can be seen in Fig. \ref{dark_spot}. Here we show the reconstruction of a star with a dark spot, whose correlation with the pristine image is $c=0.940\pm0.001$. We can compare this correlation with the correlation of the reconstructed image and a uniform disk, which is $c=0.904\pm0.001$ (lower by 30 standard deviations).\\

For both examples, we also calculate the correlation with the pristine image as a 
function of the angular scale (in $\mathrm{mas}$) of the pristine image. We then find the
degree of correlation of each reconstruction and its pristine image, and also the
degree of correlation of the reconstruction with a uniform disk.  By comparing 
these two correlation values, we can find the 
smallest feature that can be reconstructed. Below some point we
no longer distinguish between the reconstruction of the featured image and that 
of a uniform disk. We find that the smallest feature that 
can be reconstructed is close to $0.05\,\mathrm{mas}$. This can already be seen 
from the order of magnitude estimate made in section \ref{sensitivity} and a comparable 
value of $0.03\,\mathrm{mas}$ is found in section \ref{disks}. When pristine images 
have angular sizes greater than $0.8\, \mathrm{mas}$, the degree of 
correlation drops significantly due to lack of short baselines. Note that
this value is higher than what is found in section \ref{disks}. This is due to 
higher signal to noise ratio in the simulated data corresponding 
to Figures \ref{horizontal_disk} and \ref{dark_spot}. The resolution limits
discussed in turn correspond to $\sim 16\times 16$ effective resolution elements (pixels).\\

\subsection{Post-processing\label{mira}}

\begin{figure}
  \begin{center}
    \rotatebox{-90}{\includegraphics[scale=0.39]{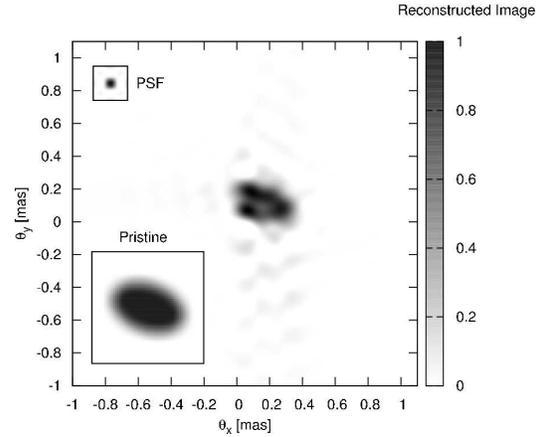}}
  \end{center}
  \caption{\label{oblate_example_c} Simulated and (raw) reconstructed oblate rotator of magnitude 3 and 10 hours
  of observation time. Reconstruction quality is lower when compared to Fig. \ref{oblate_example_a}. This is  due 
  to high frequency noise in the Hermite fit of the data, and also due to limitations of the phase reconstruction 
  algorithm. Post-processing is applied and Fig. \ref{improved_oblate} is obtained.}
\end{figure}

\begin{figure}
  \begin{center}
    \rotatebox{-90}{\includegraphics[scale=0.4]{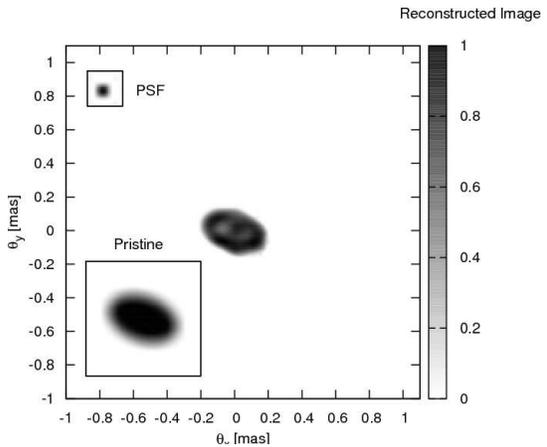}}
  \end{center}
  \caption{\label{improved_oblate} Image post-processing on Fig. \ref{oblate_example_c}. The 
    MiRA software was used to optimize on an area $0.3\,\mathrm{mas}\times0.3\,\mathrm{mas}$.}
\end{figure}

Image post-processing is performed in order to improve the raw images presented above. The type of post-processing that is currently being investigated is analogous to the data analysis performed in conventional amplitude interferometry. In  this analysis, the image is slightly modified so as to maximize the agreement between the data and the magnitude of the Fourier transform of the reconstructed image. See \citet{Thibeaut} for details. Also, one can introduce additional and very general constraints to this maximization procedure, but this is beyond the scope of this paper. Such an analysis depends strongly on the starting image, which can be provided by the analysis presented in this paper. An example of image post-processing performed on the oblate star of Fig. \ref{oblate_example_c} is shown in Fig. \ref{improved_oblate}.\\

\begin{figure}
  \begin{center}
    \rotatebox{-90}{\includegraphics[scale=0.4]{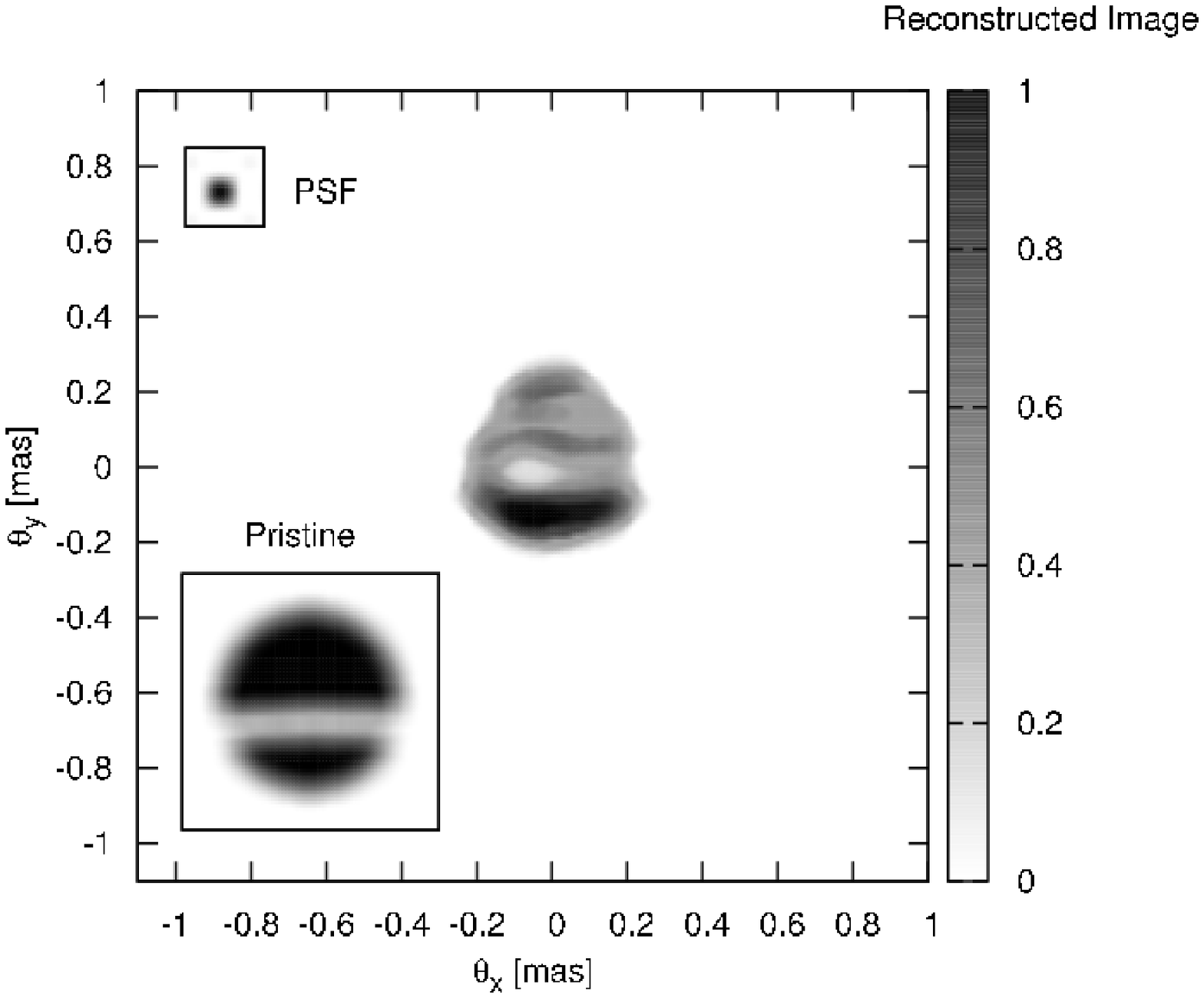}}
  \end{center}
  \caption{\label{mira_image_a} Post processed Fig. \ref{horizontal_disk}. The 
    MiRA \citep{Thibeaut} software was used to optimize on an area of $1\,\mathrm{mas}\times1\,\mathrm{mas}$.}
\end{figure}

\begin{figure}
  \begin{center}
    \rotatebox{-90}{\includegraphics[scale=0.4]{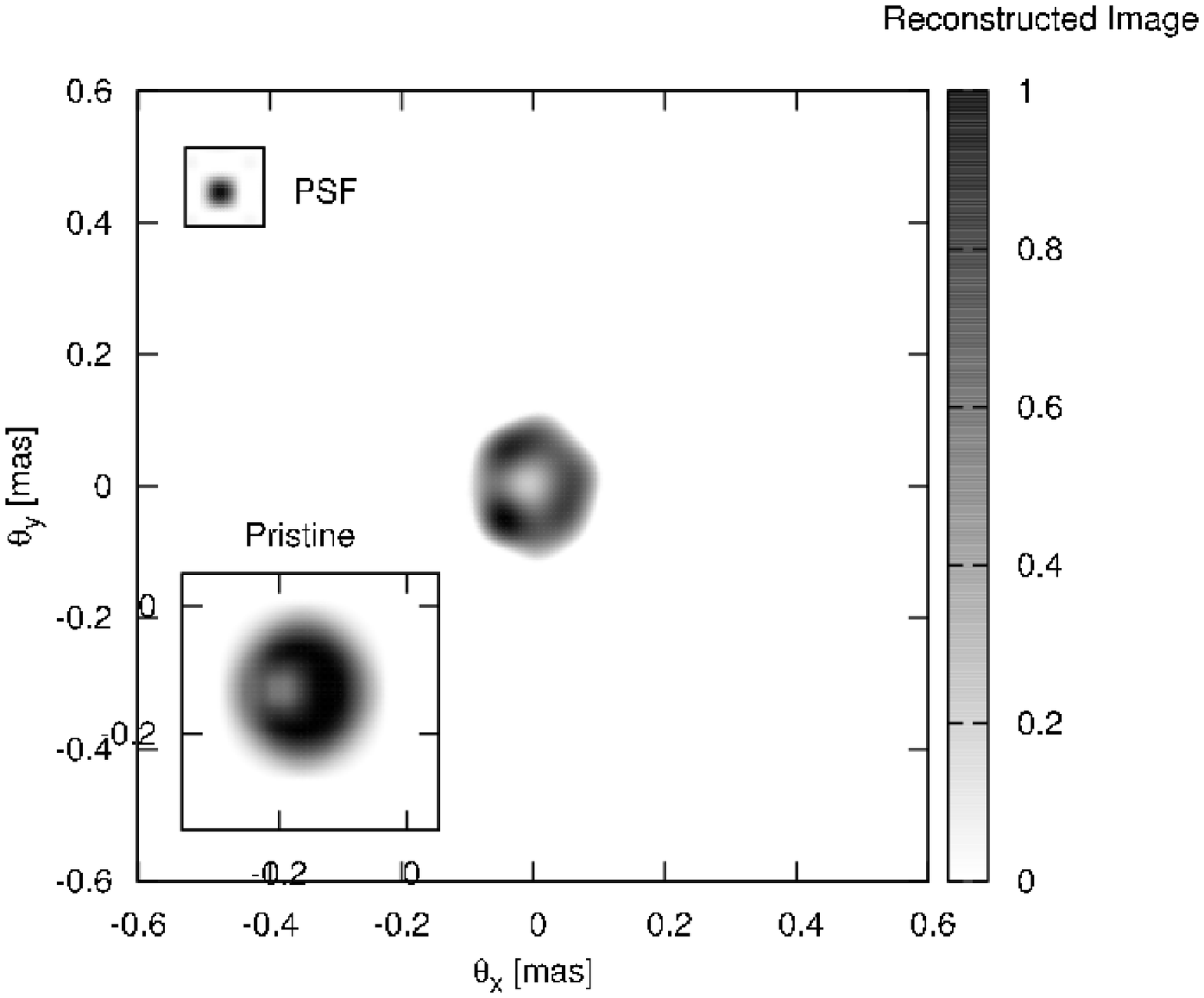}}
  \end{center}
  \caption{\label{mira_image_b} Post processed Fig. \ref{dark_spot}. The 
    MiRA software was used to optimize on an area $0.3\,\mathrm{mas}\times0.3\,\mathrm{mas}$.}
\end{figure}

We have performed this type of post-processing  on the images in Figures \ref{horizontal_disk} and \ref{dark_spot} using the MiRA software \citep{Thibeaut}, and we obtain the ones shown in Figures \ref{mira_image_a} and \ref{mira_image_b}. Aside from optimizing agreement with the data, no additional constraints are imposed. These preliminary results show the overall reduction in noise and improvement in the sharpness of the reconstructed images.  A systematic study of the improvements with image post-processing is currently being investigated \citep{Nunez}.

\section{Conclusions}

We perform a simulation study of the imaging capabilities at $400\,\mathrm{nm}$ of an IACT array consisting of 97 telescopes separated 
up to a $\mathrm{km}$. This is a preliminary design for the CTA project, expected to be operational in 2018. Our method uses a model-independent 
algorithm to recover the phase from intensity interferometric data. We test the method on images of increasing degrees
of complexity, parameterizing the pristine image, and comparing the reconstructed parameter with the 
pristine parameter. We now summarize our results and briefly discuss how fundamental stellar parameters can be constrained with the methods described above.\\

We first find that for bright disk-like stars ($M_v=6$), radii are well reconstructed from $0.03\,\mathrm{mas}$ to $0.6\,\mathrm{mas}$.
 Even though using a Cauchy-Riemann based approach to recover images
might not be the most efficient way to measure stellar radii, such a study
 starts to quantify the abilities of measuring other scale parameters in
more complicated images (oblateness, distance between binary components,
etc.). The range of angular radii that can be 
measured with a CTA-like array ($0.03\,-\,0.6\,\mathrm{mas}$) will complement existing measurements ($2\,-50\,\mathrm{mas}$) \citep{Haniff}. With the aid of
photometry, the effective temperature\footnote{Defined as $T_{eff}=\left(\frac{L}{4\pi R^2\sigma}\right)^{1/4}$, where $R$ is the radius, $L$ is the luminosity, and $\sigma$ is the Stephan-Boltzmann constant.} scale of stars within $0.05\,-\,0.5\,\mathrm{mas}$ can be extended. Multi-wavelength angular diameter measurements will also reveal the wavelength dependence of limb-darkening \citep{Aufdenberg}, and such measurements can be used to constrain energy transport models as is done in amplitude interferometry \citep{procyon}.\\

For oblate stars, similar results for extracting geometrical parameters are found. Due to the relative ease of SII to observe at short ($\sim 400\,\mathrm{nm}$) wavelengths, measuring fundamental parameters of hot B type stars is possible. B stars are particularly interesting since rapid rotation, oblateness, and mass loss are a common feature. We show that oblateness can be accurately measured, and a next step will be to quantify the capabilities of imaging realistic surface brightness distributions in hot stars. By imaging brightness (temperature) distributions, we will be able to study effects such  as limb darkening and mass loss in hot massive stars \citep{ridgway}, as well as gravity darkening \citep{von_zeipel}.\\

Binary stars are well reconstructed when one of the members is not much brighter (three times as bright) than the other, 
and when they are not too far apart ($\leq 0.75\,\mathrm{mas}$). As with amplitude interferometry, SII, along with spectroscopy, will allow the determination masses and orbital parameters in binary stars. If measured with enough precision ($\leq 2\%$) \citep{Andersen}, the determination of the mass can be used to test main sequence stellar models. An advantage of using an array such as the one used in this study, is that individual radii can be resolved. An interesting phenomena to be studied with interacting binary stars is mass transfer (e.g. \citealt{zhao}), and capabilities for imaging this phenomena can be further investigated.\\

Two examples of reconstructions of more complex images are presented, and 
demonstrate the capability to resolve features in the sub-milliarcsecond scale. Results show improvement when 
post-processing is performed. An optimization procedure improves the image reconstructions found by the methods presented above
by maximizing their agreement with the data. A more comprehensive study of reconstructions of complex images is currently being performed.

\section*{Appendix A. Cauchy-Riemann phase recovery}

Recall that since SII can only allow to measure $|\gamma|^2$, the phase information is lost, and we would like to recover it for imaging purposes, with only the magnitude information. If we denote $I(z)=R(z) e^{i\Phi(z)}$, where $z\equiv\xi+i\psi$, we obtain the following relations from the Cauchy-Riemann equations\footnote{The C-R equations can be applied because ``I'' 
is a polynomial in z.}:

\begin{eqnarray}
\frac{\partial\Phi}{\partial\psi}&=&\frac{\partial\ln{R}}{\partial\xi}\equiv \frac{\partial s}{\partial \xi}\\ 
\frac{\partial\Phi}{\partial\xi}&=&-\frac{\partial\ln{R}}{\partial\psi}\equiv -\frac{\partial s}{\partial \psi},  \label{cauchy}
\end{eqnarray}

where we have defined $s$ as the log-magnitude. Notice the relation between the magnitude and the phase. By using the Cauchy-Riemann equations we can write the log-magnitude differences along the real and imaginary axes as:

\begin{eqnarray}
\Delta s_{\xi} &=&\frac{\partial s}{\partial \xi}\Delta\xi=\frac{\partial\Phi}{\partial\psi}\Delta\xi\\ 
\Delta s_{\psi}&=&\frac{\partial s}{\partial \psi}\Delta\psi=-\frac{\partial\Phi}{\partial\xi}\Delta\psi\\
\end{eqnarray}

If the log-magnitude were available along purely the $\xi$ or the $\psi$ axes, we could solve the previous two equations for the phase.\\

However, notice that because $|z|=1$, 
we can only measure the log-magnitude on the unit circle in the complex space ($\xi,\psi$). \\

In general, we can write the log-magnitude differences along the unit circle as

\begin{eqnarray}
\Delta s_{||}&=&\frac{\partial s}{\partial \xi}\Delta\xi+\frac{\partial s}{\partial \psi}\Delta\psi \label{path0}\\
            &=&\frac{\partial\Phi}{\partial\psi}\Delta\xi-\frac{\partial\Phi}{\partial\xi}\Delta\psi \label{path}\\ 
            &=&\Delta \Phi_{\bot}. \nonumber
\end{eqnarray}

Here $\Delta \Phi_{\bot}$ corresponds to phase differences along a direction perpendicular to $\Delta s_{||}$, that is, perpendicular to the unit circle in the $\xi-\psi$ plane. We are however interested in obtaining $\Delta\Phi_{||}$, so that we can integrate along the unit circle.\\

The general form of $\Phi$ can be found by taking second derivatives in eq. (\ref{cauchy}) and thus noting that $\Phi$ is a solution of the Laplace equation in the complex plane.

\begin{equation}
\frac{\partial^2\Phi}{\partial\xi^2}+\frac{\partial^2\Phi}{\partial\psi^2}=0.
\end{equation}

The general solution of $\Phi(z)$ in polar coordinates ($\rho, \phi$) is \citep{Jackson}

\begin{equation}
\Phi(\rho,\phi)=a_0+b_0\phi+\sum_{j} \rho^j  \left( a_j \cos{j\phi}+b_j\sin{j\phi}\right) \label{solution},
\end{equation}

where terms singular at the origin ($\rho^{-j}$) have been omitted. These singular terms lead to ambiguous reconstructions including flipped images and have not been found to be essential for most reconstructions. \\

Now taking the difference of $\Phi$ along the radial direction we obtain

\begin{equation}
\Delta\Phi_{\bot}(\rho, \phi)=\sum_j \rho^j((1+\frac{\Delta\rho}{\rho})^j-1)\left( a_j \cos{j\phi}+b_j\sin{j\phi}\right).
\end{equation}

Note from eq. (\ref{path}) that the length in the complex plane associated with $\Delta s_{||}$ is $\Delta\phi=|\Delta\xi+i\Delta\psi|$, and that the length associated with $\Delta\Phi_{\bot}$ is $\Delta\rho=|\Delta\xi+i\Delta\psi|$. Now setting $\rho=1$,  $\Delta\rho=\Delta\phi$, and for simplicity of presentation, expanding for small $\Delta\phi$, we obtain

\begin{equation}
\Delta\Phi_{\bot}(\phi)=\sum_j j\Delta\phi\left( a_j \cos{j\phi}+b_j\sin{j\phi}\right).
\end{equation}

So now the coefficients $a_j$ and $b_j$ can be found using equations (\ref{path0}-\ref{path}) from the measured $\Delta s_{||}$, and thus $\Phi$ can be
found in the complex plane, with an uncertainty in $a_0$ and $b_0$. The
coefficients $a_j$ and $b_j$ can be calculated by performing the following integrals:

\begin{equation}
a_j=\frac{1}{2\pi\,j}\int_0^{2\pi} \frac{d\Phi_{\bot}}{d\phi} \cos{j\phi}\,d\phi 
\label{a}
\end{equation}

\begin{equation}
b_j=\frac{1}{2\pi\,j}\int_0^{2 \pi} \frac{d\Phi_{\bot}}{d\phi} \sin{j\phi}\,d\phi
\label{b}
\end{equation}

Note however that the previous expressions must exist, which is not the
general case. More explicitly, if the magnitude is zero, the log-magnitude
is singular. When imaging finite objects in image space, there will always be
zeros in the magnitude of the Fourier transform. In practice we are always
sample limited and nothing prevents us from calculating $a_j$ and $b_j$ approximately.

\end{document}